\documentclass[journal=jacsat,manuscript=article]{achemso}
\usepackage[version=3]{mhchem} 

\author{Jing Wang$^{1,2}$}
\author{Hong-Man Ma$^1$}
\author{Ying Liu$^{1,3}$}
\email{yliu@hebtu.edu.cn}

\affiliation
{$^1$Department of Physics and Hebei Advanced Thin Film Laboratory, Hebei Normal University,\\ Shijiazhuang 050024, Hebei, China.\\
$^2$State Key Laboratory for Superlattices and Microstructures, Institute of Semiconductors, \\Chinese Academy of Sciences, Beijing 100083, China.\\
$^3$National Key Laboratory for Materials Simulation and Design, Beijing 100083, China}

\title {Sc$_{20}$C$_{60}$: A Volleyballene}


\begin{document}

\begin{abstract}
  An exceptionally stable hollow cage containing 20 scandiums and 60 carbons was identified. This Sc$_{20}$C$_{60}$ molecular cluster has a $T_h$ point group symmetry and a volleyball-like shape, that we refer to below as "Volleyballene". Electronic structure analysis shows that the formation of delocalized $\pi$ bonds between Sc atoms and neighboring five-membered carbon rings is crucial for stabilizing the cage structure. A relatively large HOMO-LUMO gap was found. The results of vibrational frequency analysis and molecular dynamics simulations also demonstrate that this Volleyballene molecule is exceptionally stable.
\end{abstract}


Since the experimental observation of C$_{60}$\cite{1}, many very interesting structures have been proposed, such as $M_8$C$_{12}$ ($M$=Ti, V, Zr, Hf, \textit{et al})\cite{2,3}, Au$_{20}$\cite{8}, Au$_{32}$\cite{9}, Au$_{42}$\cite{10}, $M$@Si$_n$($M$=Transition Metals; $n$=14,15,16)\cite{11}, Eu@Si$_{20}$\cite{12}, Eu$_2$Si$_{30}$\cite{13}, B$_{80}$\cite{14}, and B$_{40}^{-/0}$\cite{15}. In the present work, an exceptionally stable hollow cage, Sc$_{20}$C$_{60}$ Volleyballene, has been identified by first-principles density functional theory studies.

Our first-principles calculations were performed within the framework of spin-polarized density functional theory (DFT). The exchange-correlation interaction was treated within the generalized gradient approximation (GGA) using the Perdue-Burke-Ernzerhof (PBE) exchange-correlation functionals\cite{16}. A double-numerical polarized (DNP) basis set\cite{17} was chosen to carry out the electronic structure calculation with unrestricted symmetry. For the transition metal atoms, relativistic effects in the core were included by using the DFT semi-core pseudopotentials (DSPP) \cite{18}.

\begin{figure}
 \includegraphics[scale=0.5]{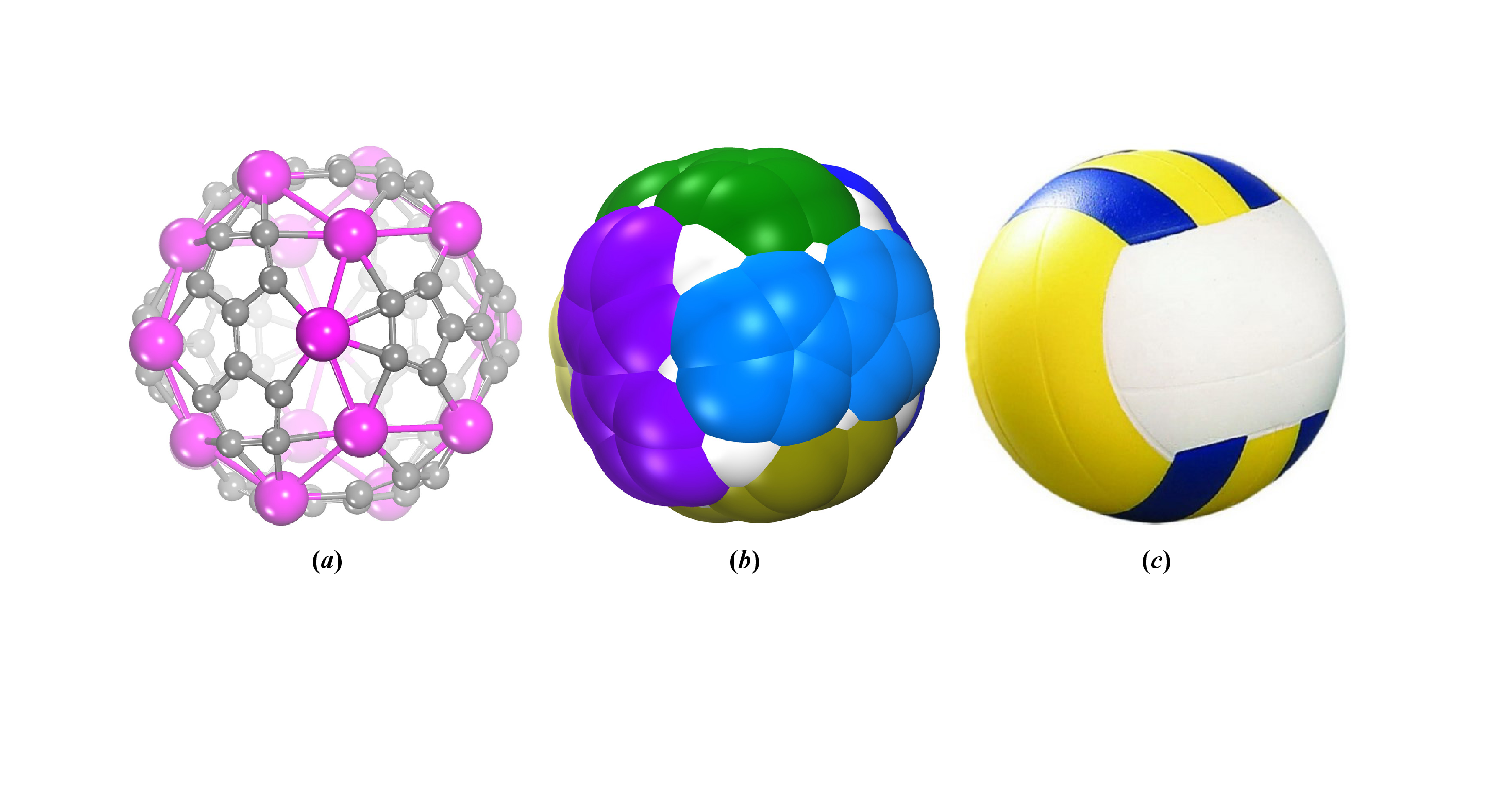}
  \caption{The configurations of the Volleyballene Sc$_{20}$C$_{60}$ viewed in ($a$) ball and stick model (Large ball: Sc atom; small ball: C atom) and ($b$) CPK style. Part ($c$) shows a volleyball.}
  \label{fgr:J.Wang.Fig1}
\end{figure}

Figure 1 shows the configuration of the Volleyballene Sc$_{20}$C$_{60}$. It is comprised of 20 scandium atoms and 60 carbons with a volleyball-like shape. This new, lowest energy Sc$_{20}$C$_{60}$ molecule had $T_h$ point group symmetry and robust stability. As may be seen from Fig. 1, it may be viewed as consisting of six Sc$_8$C$_{10}$ subunits joined together in a crisscross pattern. In this structure, there are 12 pentagonal rings made of carbon atoms (C-pentagons) and 6 octagonal rings of scandium atoms (Sc-octagons). Every group of two C-pentagons is surrounded by one Sc-octagon to give a Sc$_8$C$_{10}$ subunit.

The 20 Sc atoms link to form 12 suture lines, and the average Sc-Sc length is 3.222 {\AA}. Depending on the coordination of the Sc atoms, they may be divided into two types, Sc$^\textrm{I}$ and Sc$^{\textrm{II}}$ (see Fig. 2b). For the C-pentagons, there are three C-C double bonds and two C-C single bonds. Of the three C-C double bonds, one has a length of 1.466 {\AA} and the other two both have lengths of 1.434 {\AA}. The two C-C single bonds are both 1.446 {\AA} long. Along with a 1.463 {\AA} C-C bond connecting the two C-pentagons, the average C-C bond length is found to be 1.446 {\AA}. The average Sc-C bond length is 2.248 {\AA}.

The following aspects of the stability of the Volleyballene Sc$_{20}$C$_{60}$ were investigated: the bonding character, the binding energy, the vibrational frequencies and the molecular dynamics, the latter through ensemble simulations.

\begin{figure}
 \includegraphics[scale=0.5]{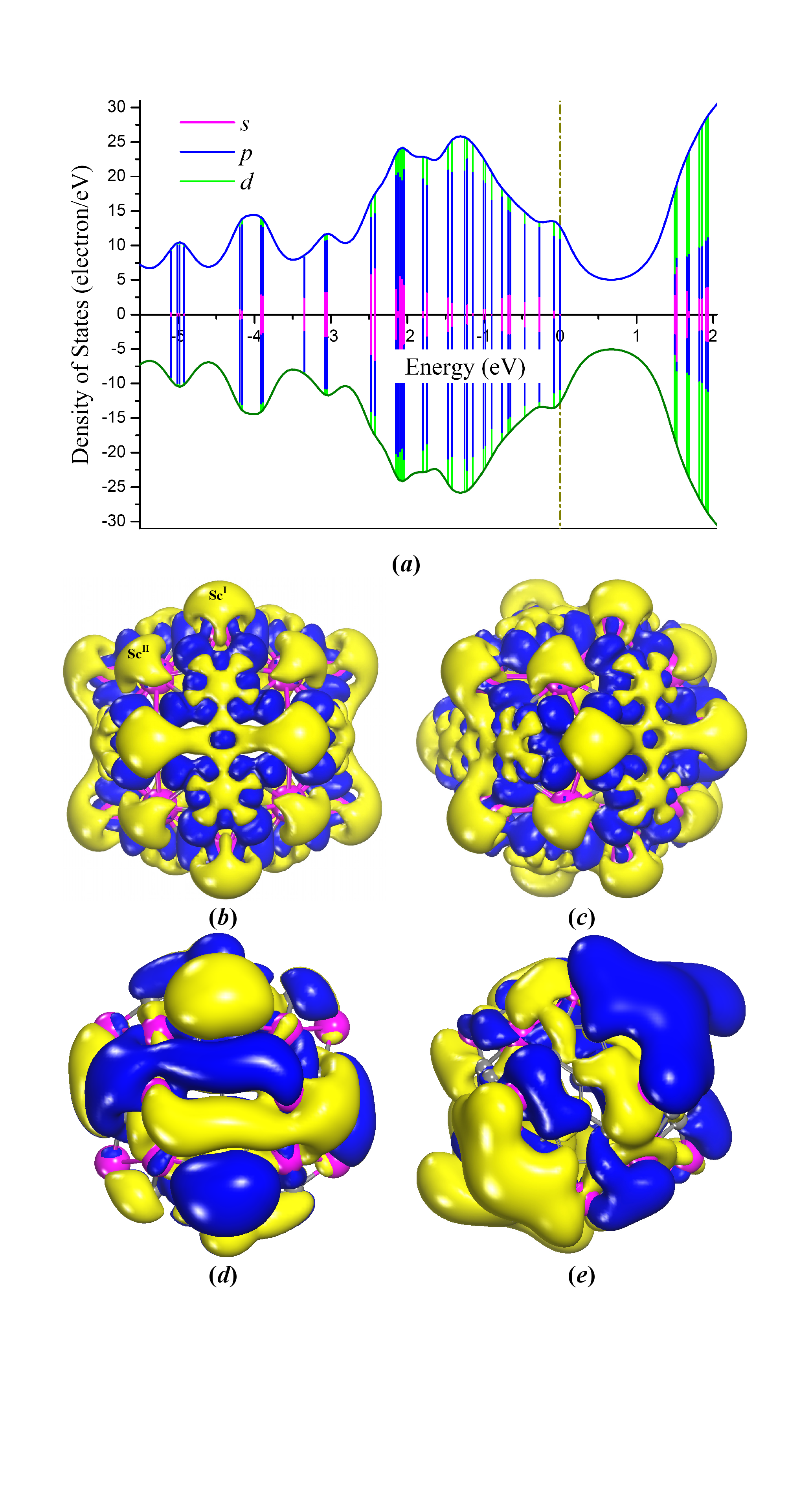}
  \caption{($a$) PDOS for the Volleyballene Sc$_{20}$C$_{60}$ , ($b$,$c$) the deformation electron density viewed from the top of Sc$_8$C$_{10}$ subunit and the suture line ($c$), and ($d$) HOMO, ($e$) LUMO orbitals for Volleyballene Sc$_{20}$C$_{60}$ . The zero of energy for the PDOS curve ($a$) is taken to be the Fermi energy, and the length of each vertical line represents the corresponding relative amplitude for each electronic orbital. The isosurface for the deformation electron density (parts $b$ and $c$) and the two orbitals (parts $d$ and $e$) are taken to be 0.030 and 0.005 $e / ${\AA}$^3$, respectively.}
  \label{fgr:J.Wang.Fig2}
\end{figure}

Firstly, the bonding character of Volleyballene Sc$_{20}$C$_{60}$ was investigated by analyzing its deformation electron density, as shown in Figs. 2$b$ and 2$c$. For the C atoms, there are obvious characteristics of $sp^2$-like hybridization, and each C atom has three $\sigma$ bonds. For the C atoms that neighbor pairs of Sc atoms, two of the three orbital lobes point towards the two neighboring C atoms, and the third lobe points towards the center point of the line joining the two Sc atoms but with a slight inclination toward Sc$^\textrm{I}$. For the other cases, the C atoms have three neighboring C atoms and form $\sigma$ bonds between all four C atoms. An outer delocalized electron is then left over for each carbon atom. The Sc atom has three outer electrons, one lying in the 3$d$ state and the other two lying in the 4$s$ state. In each Sc$_8$C$_{10}$ subunit, a delocalized $\pi$ bond forms between the two Sc$^\textrm{I}$ atoms and one C atom located at the waist position. This is crucial for stabilizing the Sc$_8$C$_{10}$ subunit. For the Sc$^{\textrm{II}}$ atoms, there is a chiral $\pi$ bond three petals pointing towards the three neighboring C-C bonds, which strengthens the link between the Sc$_8$C$_{10}$ subunits.

In order to further validate the stability of the Volleyballene Sc$_{20}$C$_{60}$, we constructed five other molecular structures with different combinations of 12 C-pentagons and 20 Sc atoms. Calculations were carried out within the same framework as described above. After energy minimization, it was found that the optimized structures all showed relatively large deformations and did not retain their original topologies. However, closer observation of these configurations indicated that although the overall configurations were not viable, one or more Sc$_8$C$_{10}$ subunit usually appeared, which further suggests that the Sc$_8$C$_{10}$ subunit is very stable. Figure 3 shows three of the cage-like geometries examined before and after optimizations. Beneath each isomer is listed the relative binding energy ($\Delta E_b$) with respect to the Volleyballene. Clearly, from the energy point of view, the Volleyballene (shown in Fig. 1) was the lowest energy structure.

\begin{figure}
 \includegraphics[scale=0.4]{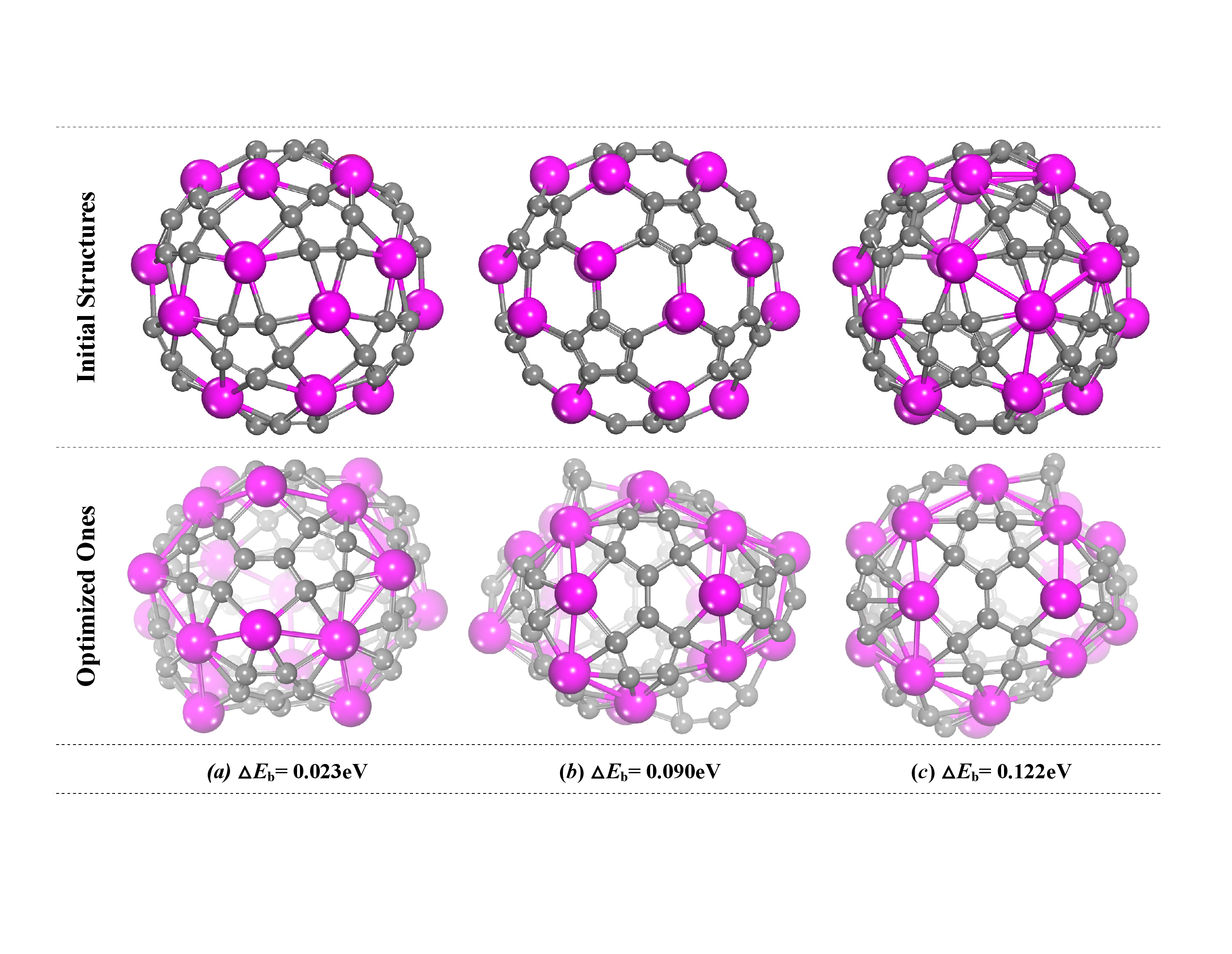}
  \caption{Three kinds of typical initial and optimized configurations of Sc$_{20}$C$_{60}$  clusters. Beneath each isomer is listed the relative binding energy ($\Delta E_b$) with respect to Volleyballene Sc$_{20}$C$_{60}$ . Key: large ball, Sc atom; small ball, C atom.}
  \label{fgr:J.Wang.Fig3}
\end{figure}

Next, a vibrational frequency analysis was carried out for the lowest energy Volleyballene, Sc$_{20}$C$_{60}$. There were no imaginary frequencies and the two highest intensity frequencies were found to be 468.9 and 472.3 $cm^{-1}$. Figure 4 gives the calculated Raman spectrum with a 300 K temperature and 488.0 nm incident light, in order to simulate a realistic Raman spectrum that can be compared to experimental results. The calculated results show that the Volleyballene Sc$_{20}$C$_{60}$ is kinetically the most stable isomer.

\begin{figure}
 \includegraphics[scale=0.45]{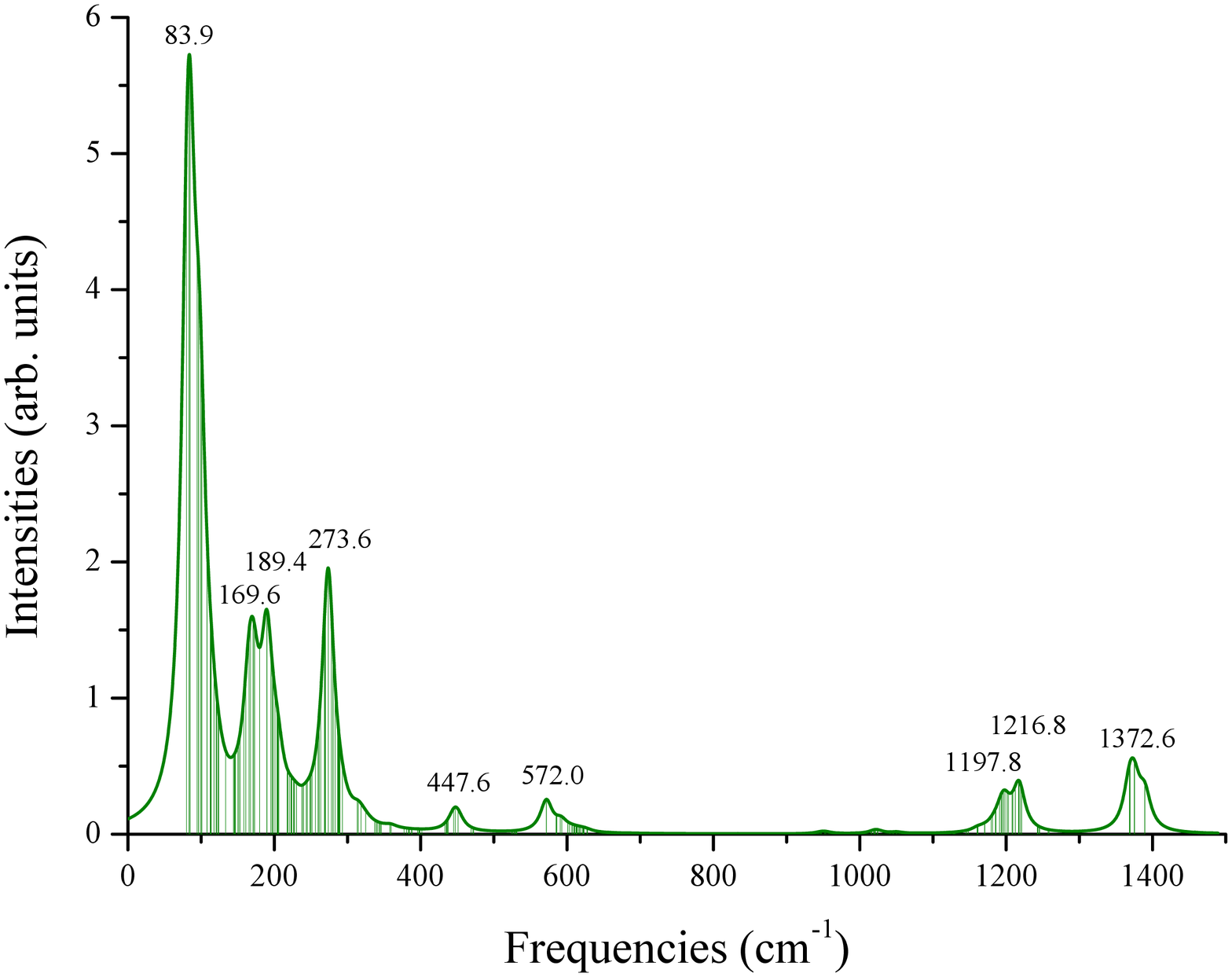}
  \caption{The simulate Raman spectrum for the Volleyballene Sc$_{20}$C$_{60}$ with a 300 K temperature and 488.0 nm incident light. The Lorentzian smearing is set to be 20.00 cm$^{-1}$. The labels present the frequencies corresponding to the peaks of the intensities.}
  \label{fgr:J.Wang.Fig4}
\end{figure}

In addition, ab initio molecular dynamics simulations with NVE and NVT canonical ensembles were carried out to test the thermodynamic stability of the Sc$_{20}$C$_{60}$ Volleyballene. For the NVE ensemble, the total simulation time was set to be 2.0 $ps$ with a time step of 1.0 $fs$ at initial temperatures of 1000, 1400, 2000, and 2400 $K$. The results of the NVE simulations showed that the structure of Volleyballene was not disrupted over the course of a 2.0 $ps$ dynamical simulation at an initial temperature of 2400 K, equal to a $\sim$1200 K effective temperature. For the NVT ensemble, the Gaussian thermostat\cite{19} was chosen, and the total run time was set to be 1.0 $ps$ with a time step of 1.0 $fs$ at temperatures of $T$ = 800 and 1000 $K$. The NVT dynamical simulations indicated that the Volleyballene Sc$_{20}$C$_{60}$ also retained its original topological structure up to a temperature of 1000 $K$. These results indicate that the Sc$_{20}$C$_{60}$ Volleyballene has good thermodynamic stability.

Considering further the question of stability, it is natural to explore the electronic structure. To this end, we calculated the partial density of states (PDOS) and frontier molecular orbitals, including the highest occupied molecular orbital (HOMO) and the lowest unoccupied molecular orbital (LUMO) of Volleyballene as shown in Figures 2$a$, 2$d$, and 2$e$. From the contours of the HOMO and LUMO orbitals, it may be seen that the HOMO orbitals are mostly localized on the C atoms and Sc$^\textrm{I}$ atoms. There is also obvious $p$-$d$ hybridization. As for the case of the LUMO, the energy level features show that the LUMO orbital is doubly degenerate. On the Sc atom, there are obvious $d$-orbital characteristics, with $d_{z^2}$-like orbitals for Sc$^{\textrm{II}}$ atoms and other $d$-like orbitals for Sc$^\textrm{I}$. The LUMO orbital hybridization is predominantly $sp$-$d$ hybridization. Close examination of the PDOS (see Fig. 2a) further confirms the hybridization characteristics of the HOMO and LUMO orbitals. All these results demonstrate that the hybridization between Sc $d$ orbitals and C $p$ orbitals is essential for stabilizing the cage structure. For the Volleyballene Sc$_{20}$C$_{60}$, a relatively large HOMO-LUMO gap of 1.471 eV was observed, mainly due to the energy of the d atomic orbitals being much lower than that of the $p$ orbitals. The Volleyballene Sc$_{20}$C$_{60}$ should therefore be an exotic buckyballene variant with exceedingly high chemical stability.

\begin{acknowledgement}
The authors thank the science writers' reports\cite{+1} on this work and Dr. N. E. Davison for his help with the language. This work was supported by the National Natural Science Foundation of China (Grant Nos. 11274089 and 11304076), the Natural Science Foundation of Hebei Province (Grant Nos. A2012205066 and A2012205069), and the Science Foundation of Hebei Education Award for Distinguished Young Scholars (Grant No. YQ2013008). We also acknowledge partially financial support from the 973 Project in China under Grant No. 2011CB606401.
\end{acknowledgement}

\end{document}